\documentclass[pra,aps,twocolumn]{revtex4}

\usepackage{graphics,epsfig ,color,graphicx} 

\begin{document}
\date{\today}
\title[Grading uncertainty]{A priori analysis:  an application to the estimate of the uncertainty in course grades}
\author{G.L. Lippi$^{1,2}$ \\}
\address{
$^1$ Institut Non Lin\'eaire de Nice, Universit\'e de Nice Sophia
Antipolis\\
$^2$ CNRS, UMR 7335\\
1361 Route des Lucioles, F-06560 Valbonne, France\\
Gian-Luca.Lippi@inln.cnrs.fr}

\begin{abstract}
The {\it a priori analysis} ({\it APA}) is discussed as a tool to assess the reliability of grades in standard curricular courses.  This unusual, but striking application is presented when teaching the section on data treatment of a Laboratory Course to illustrate the characteristics of the {\it APA} and its potential for widespread use, beyond the traditional Physics Curriculum.  The conditions necessary for this kind of analysis are discussed, the general framework is set out and a specific example is given to illustrate its various aspects.  Students are often struck by this unusual application and are more apt to remember the {\it APA}.  Instructors may also benefit from some of the gathered information, as discussed in the paper.
\end{abstract}

\maketitle

\section{Introduction}\label{intro}

Teaching statistical data treatment techniques, generally done within the framework of a laboratory course in the Physics Curriculum, is a task that is known to many of us as an unrewarding one.  On the one hand, the topics to be treated need attention to detail and to the basic assumptions which ensure the validity of the whole analysis.  On the other hand, students often find the topic boring and unappealing.  Nonetheless, it is a job that needs to be accomplished, in the same way one has to learn many other basic techniques indispensable for the conduct of one's work.  Thus, it is always helpful to try and find ways of rendering the material more appealing to students, for example by using some unusual and unexpected applications of data treatment.

One such example, which infallibly attracts students' attention, is the use of the {\it A Priori Analysis} ({\it APA}) to estimate the uncertainty on the grade which they receive in the same laboratory course where I present this material.  Besides attracting the students' attention, this example never fails to arouse some curiosity -- often disguised -- since the concept of a grade not being given with perfect certainty appears to surprise a number of students (and perhaps unsettle some of them a little bit).

Thus, in addition to giving students an illustration of the {\it APA} which brings the message very close to home, this application carries a triple weight:  1. giving a practical implementation of a concept which may otherwise be relegated to the category of techniques to be set aside (and forgot)\footnote{Not too many students in a class will be confronted with the needs of a true {\it APA} in the context of their future careers.}; 2. introducing the pedagogically important concept that grades, like everything else in real life, are affected by an intrinsic uncertainty; 3. showing that the tools that are learnt within the Physics Curriculum have an application to real life.

Instructors may also find that this application of the {\it APA} offers some useful information, as illustrated in the course of the paper and discussed in more detail in the conclusions.

\section{A Priori Analysis}\label{apa}

Although not always included among the experimental analysis techniques taught in the basic curriculum, the {\it APA} is a powerful method which makes it possible to identify the different sources of uncertainty and quantify their influence on the outcome of an experiment.  Two applications of the {\it APA} are obvious:  1. estimating the size of errors which can be expected {\bf before} performing an experiment -- particularly interesting for long and complex or expensive experiments -- and 2. evaluating the influence of the individual error sources, thus permitting the identification and/or removal of the strongest uncertainty contributions.  In this sense in Physics one could say that the {\it APA} is most useful in experiment design and/or performance evaluation.~\footnote{It is also useful to check whether the errors obtained are consistent with those expected, ths enabling an independent test of the error size and, possibly, the detection of mistakes (or systematic errors).  This is often beyond the reach of a student lab, though.}  However, in addition to these two more immediate applications, the {\it APA} becomes a valuable tool of error assessment whenever the {\it a posteriori} analysis -- through repetition of the experiment -- is not possible\footnote{A good example within the physical sciences can be given when considering working on existing datasets -- e.g., old climatologic or meteorological data, where only measurements without uncertainties may have been registered.}.

Since the concept of repetition may be arguable in the example that I will discuss (tests can be repeated, and multiple tests are regularly administered in a course), it is useful to recall the constraints which need to be fulfilled for the {\it a posteriori} analysis to hold~\cite{Taylor1982}.  Indeed, since mean and standard deviations are estimated through the repetition of the measurement, {\it each outcome must be statistically independent from the previous ones} (i.e., the fluctuations from one measurement to the other be random).  If this condition is not fulfilled, the estimated mean and standard deviation do not hold and the results lose significance.

The statistical independence hypothesis certainly does not hold for at least two reasons:
\begin{itemize}\itemsep0cm
\item[1.] it is virtually impossible to devise tests which are perfectly equivalent, thus the {\it experimental conditions} \underline{are not constant};
\item[2.] one cannot repeat the {\it experiment} by having students take repeated (equivalent) tests, since the (desirable, and normally observed) progressive improvement accompanying successive tests would skew the results (particularly the standard deviation).  Hence, the condition of statistical independence is clearly violated.
\end{itemize}

While one could think of finding ways of compensating for the difficulty presented in (1.) by taking large ensemble averages (i.e., repeated tests for each student) which could smooth out the differences if the tests are sufficiently well constructed, the obstacle presented by point (2.) is unsurmountable and even contradicts the possible solution just envisaged for point (1.).  Indeed, in addition to obtaining meaningless results with test repetition, one cannot even think of replacing multiple tests with ensemble averages -- on the class size -- taken on a single test, since their indicators (average and standard deviation) cannot give any information on the grade of each individual student!  The variability in level for an entire class is tipically {\bf much} wider than the accuracy with which we can estimate the grade for an individual, since the former represents the excursion in achievement due to different levels of individual ability, assiduity, performance, engagement, etc.  Indeed, if this were not the case individual grades would be meaningless.

The {\it a priori} estimate of the uncertainty is therefore an interesting indicator which, far from providing \underline{the} correct uncertainty, gives for it at least a reasonable estimate.  As always true for {\it a priori} uncertainties, the quality of the final outcome -- at the end of the analysis -- is strictly related to the reliability of the guesses for the uncertainty of each test component.  The appropriation of this concept is pedagogically very important, as it teaches the student to critically analyze the problem and shows that within the framework of an {\it APA} a critical eye and repeated tests (with different estimates for the different error components) play a major role in the process.  Testing various estimated initial uncertainties is all the more useful the more numerous the tests which compose the final grade (although the case I discuss in detail turns out to be very simple).  As such, lab courses may be the most interesting examples, but the technique is applicable to any kind of course.

\section{Mathematical formulation}

One can generally formulate the problem as follows.
An ensemble of $N$ tests of different nature -- lab report, written test, final exam, etc. --, with individual grades $Gj$, combine with weight coefficients $w_j$ to give the global grade $G$:
\begin{equation}
\label{gradedef}
G = \sum_{j = 1}^N w_j G_j \, . 
\end{equation}

Each category of test may itself be subdivided into different subtests and result therefore from an average over homogeneous grades:
\begin{equation}
G_j = \sum_{k = 1}^M \frac{G_{j,k}}{M} \, ,
\end{equation}
where $M$ is the number of homogeneous tests in each category.  A concrete example\footnote{This is the structure of grades of the laboratory course which I use as an example for the students.} can illustrate the structure of the grades more easily.  Suppose that the grade be composed of:
\begin{itemize}\itemsep0cm
\item[1.] Work performance during the lab session; 
\item[2.] Quality of the reporting in the labbook; 
\item[3.] Evaluation of the lab report; 
\item[4.] Final exam;
\end{itemize}
thus of $N=4$ different kinds of tests.  Each category of test may contain repetitions of individual tests.  For instance, a student will do several, $M$, experiments and therefore will have at least $M$ notes in category 1.

We first define the uncertainty in the homogeneous test category as 
\begin{equation}
\label{gen-k-unc}
\sigma_{G_j} = \frac{1}{M} \sqrt{\sum_{k=1}^M \sigma_{j,k}^2} \, ,
\end{equation}
where $\sigma_{j,k}$ represents the uncertainty estimated (in the {\it APA}) for each individual test within a category.  Thus, equation~\ref{gen-k-unc} provides the general expression for the {\it a priori} estimate of the uncertainty in each grade category\footnote{This can be useful when one test from an ensemble proves more difficult to grade, or exhibits larger fluctuations in students' performance (e.g., due to a harder problem set).}.

In most cases, however, the uncertainty can be estimated to be the same for all $k$ tests of a certain category $j$ (e.g., homework, lab report, etc.).  In such a case the uncertainty, equation~\ref{gen-k-unc}, simplifies to become~\cite{Taylor1982}
\begin{eqnarray}
\label{simple-k-unc}
\sigma_{G_j} & = & \frac{1}{M} \sqrt{ \sum_{k=1}^M \sigma_j^2} \, , \\
\label{simple-ind}
& = & \frac{\sigma_j}{\sqrt{M}} \, .
\end{eqnarray}

In order to obtain the uncertainty on the final grade, it suffices to propagate the individual uncertainties $\sigma_{G_j}$ through the general definition, equation~\ref{gradedef}, to obtain~\cite{Taylor1982}
\begin{eqnarray}
\label{gen-fin}
\sigma_G & = & \sqrt{\sum_{j=1}^N \left( \frac{\partial G}{\partial G_j} \right)^2 \sigma_{G_j}^2} \, , \\
\label{simple-fin}
& = & \sqrt{ \sum_{j=1}^N \left( w_j^2 \frac{\sigma_j^2}{M} \right) } \, ,
\end{eqnarray}
where the former expression is general and the latter applies to the case of equal estimated uncertainties within a test category (cf. equation~\ref{simple-k-unc}).

\section{Estimating the a priori uncertainties}\label{apu}

The most interesting, and challenging, part of the work comes when one has to determine reasonable estimates for the uncertainties to be attributed to each individual type of test\footnote{The nature of the problem is the same if different uncertainties are attributed to the individual test, as previously mentioned.}.  
For clarity, I will proceed with a concrete example:  the laboratory course in which this material is presented.  The structure of the course is such that students are evaluated in four different categories (cf. table~\ref{grades}).

\begin{table}
\caption{
The {\it Repetitions} column corresponds to the number of tests in the corresponding category.  Different lab sessions (4 in this example) lead to one report, thus the number of lab reports is four times smaller than the number of grades in the participation or labbook sections.
}
\label{grades}
\begin{tabular}{c c c c c}\hline
Kind of test & Kind of evaluation & label ($j$) & $w_j$ & Repetitions
\\  \hline
Participation & individual & $p$ & 0.1 & 12 \\
Labbook & collective & $l$ & 0.25 & 12 \\
Report & collective & $r$ & 0.25 & 3  \\
Oral & individual & $o$ & 0.4 & 1 \\
\hline 
\end{tabular}
\end{table}

Assuming that the uncertainties be homogeneous for each category of test, we apply equations~\ref{simple-ind} and \ref{simple-fin} and therefore need to estimate the values of $\sigma_j$.  The estimates are given in table~\ref{estimates} (second column) and are based on the following considerations (items labelled according to the  test category, cf. tables~\ref{grades} and~\ref{estimates} -- for a description of French grades look at table's~\ref{estimates} caption):  
\begin{itemize}\itemsep0cm
\item[$p$] assuming $\sigma_{p} = 1$ amounts to saying that an error in grading by $\pm 3$ units has a probability of occurrence $P < 0.003 $\footnote{Gaussian-distributed errors are assumed here~\cite{Taylor1982}.}.  Translated in percentage $ 3 \widetilde{\sigma}_p = 0.15$, which is quite a large interval.  Such a large error bar is introduced on the basis of the nature of the evaluation:  different lab monitors give an estimate of the performance of each individual student -- working in a small group (typically two or three) -- by observing their work, discussing with the group and asking occasional questions.  Each monitor is required to follow several groups (typically between four and six) and differences in evaluation among monitors, as well as fluctuations for a same monitor due to variable working conditions, are unavoidable.
\item[$l$] $\sigma_l = 0.5$ amounts to assigning $\pm 1.5$ points to the uncertainty with probability $P > 0.997$ of the true grade falling within the interval.  In percentage this amounts to $3 \widetilde{\sigma}_l = 0.075$.  The variability in the evaluation is estimated to be lower than for p-tests due to the fact that labbooks, as a written document, can be more reliably evaluated.  The estimated uncertainty could be smaller if all grading were done by one and the same person (not the case in this context).  The size of $\sigma_l$ is therefore chosen to reflect the added variability coming from an ensemble of graders.
\item[$r$] We assign the same error estimates to this test as those chosen for $\sigma_l$ for the  reasons exposed in the preceding point.
\item[$o$] This kind of test requires a closer look at its details.  Being an oral examination -- even though conducted by a panel of (at least) three examiners -- it is somewhat more susceptible to fluctuations (in the questions, their evaluation, and in the student's reactions).  Therefore, we assign to it a value of $\sigma_o = 0.7$ which amounts to considering a full 95\% confidence interval\footnote{It is interesting here to use a different way of evaluating the confidence interval.  Indeed, contrary to common practice in Physics, in most applied sciences -- e.g. Risk Assessment, Geosciences, etc. -- the standard error bar associated with any given quantitative estimate is $\pm 2 \sigma$, rather than $\pm \sigma$ by virtue of its larger confidence level ($\tilde 95\%$), more useful for practical purposes.  It is therefore instructive to use this example to present this aspect of uncertainty estimates to physics students.} ($\pm 2 \sigma_o$)~\cite{Taylor1982} to a spread of (approximately) three points.  Translated into percentages, $\widetilde{\sigma}_o = 0.035$ (i.e. $3 \widetilde{\sigma}_o = 0.105$).  In other words, we expect the probability of a grade outside this interval to be below $5\%$.
\end{itemize}

\begin{table}
\caption{
The French University Grading System (FUGS) attributes the grades in $x/20$ (passing grade $x=10$), where $x$ represents the grade attributed to the test.  The numerical estimates are given in absolute values, i.e., in FUGS units, but -- in order to improve readability -- are also repeated in percentage.  The latter are identified by the corresponding quantities marked by a tilde $\widetilde{v}$ ($v$ being the generic variable).  The conversion gives rise to a resuilt with an excess of digits for some grade categories (kept here to be consistent with the absolute estimates, used for the calculations).
}
\label{estimates}
\begin{tabular}{c c c c c}\hline 
Label & $\sigma_j$ & $\sigma_{G_j}$ & $\widetilde{\sigma}_j$ & $\widetilde{\sigma}_{G_j}$ 
\\  \hline 
$p$ & 1 & 0.4 & 0.05 & 0.02 \\
$l$ & 0.5 & 0.2 & 0.025 & 0.01 \\
$r$ & 0.5 & 0.2 & 0.025 & 0.01 \\
$o$ & 0.7 & 0.7 & 0.035 & 0.035 \\
\hline 
\end{tabular}
\end{table}

The propagation of the {\it a priori} uncertainties on each individual grade for each kind of test follows equation \ref{simple-ind} and produces the values of $\sigma_{G_j}$ given in Table~\ref{estimates}.  Notice that $M=1$ for the oral test (label $o$), therefore no uncertainty improvement ensues for this grade.

Computing the propagation of the {\it a priori} uncertainty on the final grade, equation~\ref{simple-fin}, we obtain 
\begin{eqnarray}
\label{numerror}
\sigma_G & = & 0.3 \, {\rm points} \, , \\
& = & 0.015 \, {\rm (in \, percent)}
\end{eqnarray}
which amounts to a $\Delta G \simeq 0.5$ points (or $\widetilde{\Delta G} \simeq 0.025$) with probability $P \simeq 0.9$\footnote[1]{From~\cite{Dwight1961} we obtain that $P (x) = 0.9$ when $x \simeq 1.645$.  Thus, computing $\Delta G = 1.645 \times \sigma_G$ (equation~\ref{numerror}) we obtain $\Delta G \simeq 0.4935 \approx 0.5$, as in the text.} of obtaining the actual grade within this interval.

We thus conclude that the {\it a priori} estimate for the grade each student receives in this course is $\pm 0.5$ points (or $2.5\%$) with a confidence level of $90\%$.

\section{Discussion}

\begin{table}
\caption{
Individual contributions to the final grade uncertainty, estimated for each grade category according to equation~\ref{gen-fin} $\left( \frac{\partial G}{\partial G_j} = w_j \right)$.  We remark that only the last term in the table is significant.
}
\label{numerical}
\begin{tabular}{c c c c c}\hline 
 & $p$ & $l$ & $r$ & $o$ 
\\  \hline 
$w_j^2 \sigma_{G_j}^2$ & 0.0008 & 0.001 & 0.005 & 0.08 \\
\hline 
\end{tabular}
\end{table}

Aside from the numerical result just obtained, equation~\ref{numerror}, it is very instructive to look at the details of the contributions which compose the final value $\sigma_G$.  Table \ref{numerical} provides the breakup of the various contributions, where we notice that the smallest one comes from the $p$ component.
We immediately recognize that the influence on the final uncertainty coming from the participation grade ($p$) is entirely negligible (by two orders of magnitude), in spite of its intrinsic variability and of the large {\it a priori} uncertainty we have consequently assigned to it.  This results from the combined effect of the larger number of tests in this category ($M_p = 12$) and of the small weight assigned to this category ($w_p = 0.1$, cf. table~\ref{grades}).

The relative contributions of the labbook ($\sigma_l$) and report ($\sigma_r$) uncertainties are different in table~\ref{numerical} due to their different number of samples ($M_l = 12$ and $M_r = 3$), which reduce by $M^{-\frac{1}{2}}$ their uncertainty (equation~\ref{simple-ind}).  Overall, however, even the weighted contribution coming from $G_r$ is negligible -- by more than one order of magnitude -- when compared to that of the oral exam.  We thus conclude that only the uncertainty on the latter matters in the determination of the uncertainty on the final grade, owind to the larger size of $\sigma_o$, the single event ($M_o=1$), and especially its dominant weight ($w_0 = 0.4$, table~\ref{grades}).

One should not confuse the influence of each grade category on the final outcome with the dominance of the uncertainty of the oral test on the overall uncertainty.  Each grade category contributes, proportionally to its weight, to the course grade, but the confidence interval is determined exclusively by the oral test, all other forms of grading providing a much more ``accurate" evaluation.

This result has the following implications:
\begin{itemize}\itemsep0cm
\item[a.] given the very sizeable difference it error contribution (table~\ref{numerical}) modifying the estimates of the {\it a priori} uncertainties which we have assigned to the various kind of tests (except for the oral test) will not influence the size of the uncertainty.  Thus, except for the oral test, we realize a posteriori that our careful estimates in the preceding section do not hold any relevance;
\item[b.] given that the only dependence of the estimated {\it a priori} final uncertainty has a linear dependence onto the estimated error assigned to the oral test, we know that modyfing the latter linearly translates onto the reliability of the global grade (weighted by $w_o$);
\item[c.] there is no need to worry about the reliability of the grades for the first three kinds of tests, i.e. about the variability originating from the involvement of several monitors in the various grading steps.
\end{itemize}

The last point is important for instructors who may worry that, in particular for the participation grade, the instrinsic variability due to multiple evaluators, and the ensuing point spread, may distort the reliability of the course's global grade.  This also means that one can confidently introduce different measures of evaluation -- in particular some which give the benefit of an immediate return to the students, such as the participation grade -- without risking a substantial impact onto the final grade.

One final point:  the {\it a priori} estimate of the uncertainty on the final grade also gives a measure of the precision with which the latter can be given.  In the specific case of the example used, a good discretization is $\Delta G$, i.e., using a scale in points with integer and half-integer values (or 2.5\% in relative precision).  This can be used to explain to students what is a reasonable scale in grade spacing.  Of course, the actual value will depend on the structure of the course and on the number of test categories (and of test number in each category).

\section{Conclusions}

The simple, but striking, application of the {\it APA} to course grades illustrates quite effectively the intrinsic nature of this kind of analysis and its main features.  It clearly shows the technique's importance in all those situations where measurements cannot be repeated to obtain {\it a posteriori} error estimates, and the power of its predictions.  At the same time, the analysis has shown the need for a careful assessment of the individual error sources to be assigned to the primary, {\it measured} quantities, and the futility of part of the work, rendered irrelevant by the intrinsic structure of the analyzed quantities (composition of the grade and of its uncertainty).  Students are often taken aback both by the fact that an aspect of their curriculum can be analyzed in detail with techniques {\it seemingly exclusively devised} for lab experiment analysis, and by the information which can be gathered by this analysis.  This example should also serve as an encouragement for testing the application of data treatment techniques to real-life everyday's problems.

As a bonus, we have shown that instructors may gather precious information on the uncertainty which affects their grades and on the confidence level of each grade component, while gaining some freedom in experimenting with innovative ways of introducing partial grades.  The latter can be beneficial to giving students early and welcome feedback, while ensuring that the reliability of the course grade is not affected by evaluation components which are more prone to larger fluctuations.  This quantitative analysis, though partly subjective (in the assignment of the {\it a priori} error components) may also be helpful in arguing in favour of the introduction of complementary grade parts in the discussions with skeptical collegues or Department Directors.

I am grateful to all the students (in excess of five hundred) who, having taken this course over the last decade, have stimulated the development of new ideas and examples.

\end{document}